\documentclass[graybox]{svmult}

\usepackage{mathptmx}       
\usepackage{helvet}         
\usepackage{courier}        
\usepackage{type1cm}        

\usepackage{color}
\usepackage{amssymb}                            
\usepackage{makeidx}         
\usepackage{graphicx}        
\usepackage{multicol}        
\usepackage[bottom]{footmisc}

\usepackage{lipsum}

\usepackage{amsmath}

\newcommand{\su}{\mathsf{SU}}

\newcommand{\quotes}[1]{``{#1}''}

\graphicspath{{graphics/}}

\makeindex             

\begin{document}

\title*{Joint Contour Net analysis of lattice QCD data}
\author{Dean P. Thomas, Rita Borgo, Hamish Carr and Simon Hands}

\institute{Dean P. Thomas \at Departments of Computer Science and Physics, College of Science, Swansea University, Swansea, UK \email{798295@swansea.ac.uk}
\and Rita Borgo \at Informatics Department, King’s College London, London, UK \email{rita.borgo@kcl.ac.uk}
\and Hamish Carr \at School of Computing, University of Leeds, Leeds, UK \email{h.carr@leeds.ac.uk}
\and Simon Hands \at Department of Physics, College of Science, Swansea University, Swansea, UK \email{s.j.hands@swansea.ac.uk}}
%
%
\maketitle

\abstract*{Lattice Quantum Chromodynamics (QCD) is an approach used by theoretical physicists to model the strong nuclear force. This works at the subnuclear scale to bind quarks together into hadrons including the proton and neutron. One of the long term goals in lattice QCD is to produce a phase diagram of QCD matter as the thermodynamic control parameters temperature and baryon chemical potential are varied. The ability to predict critical points in the phase diagram, known as phase transitions, is one of the on-going challenges faced by domain scientists.  In this work we consider how multivariate topological visualisation techniques can be applied to simulation data to help domain scientists predict the location of phase transitions. In the process it is intended that applying these techniques to lattice QCD will strengthen the interpretation of output from multivariate topological algorithms, including the joint contour net. Lattice QCD presents an interesting opportunity for using these techniques as it offers a rich array of interacting scalar fields for analysis; however, it also presents unique challenges due to its reliance on quantum mechanics to interpret the data.}

\abstract{Lattice Quantum Chromodynamics (QCD) is an approach used by theoretical physicists to model the strong nuclear force. This works at the subnuclear scale to bind quarks together into hadrons including the proton and neutron. One of the long term goals in lattice QCD is to produce a phase diagram of QCD matter as the thermodynamic control parameters temperature and baryon chemical potential are varied. The ability to predict critical points in the phase diagram, known as phase transitions, is one of the on-going challenges faced by domain scientists.	In this work we consider how multivariate topological visualisation techniques can be applied to simulation data to help domain scientists predict the location of phase transitions. In the process it is intended that applying these techniques to lattice QCD will strengthen the interpretation of output from multivariate topological algorithms, including the joint contour net. Lattice QCD presents an interesting opportunity for using these techniques as it offers a rich array of interacting scalar fields for analysis; however, it also presents unique challenges due to its reliance on quantum mechanics to interpret the data.}

\section{Introduction}
\label{sec::introduction}


Multivariate topology can assist scientists in various domains understand correlations between different fields defined upon common sampling points.  Several approaches exist including Jacobi Sets, Reeb graph comparison, Reeb spaces, and range tessellation.  In this paper we concentrate on the Joint Contour Net (JCN)~\cite{Carr2013} which builds upon existing theoretical and practical aspects of the Reeb graph to represent the Reeb space in a discrete graph based format.  The Reeb space~\cite{edelsbrunner2008reeb} addresses the relation between multiple sampled fields by contracting multivariate contours to singular points.


In this work we use the JCN, previously used to analyse nuclear scission~\cite{schunck2014description, schunck2015description} and hurricane datasets~\cite{geng2014visual}, to analyse lattice Quantum Chromodynamics (lattice QCD) simulations.  In order to carry out this process, the data is first visually examined to form a link between the graph-like structure of JCN with a geometric segmentation of the multi-field.  These observations enable the formulation of a number of quantitative properties to be extracted from the bivariate topology, providing a comparison of multiple data sets with different simulation parameters.  In doing so we make the following contributions:
\begin{itemize}
	\item Extend the use of the multivariate topological techniques to a new scientific domain, lattice QCD
	\item Show how analysis of the JCN using non-visual techniques can allow it to be used on large, complex data sets that are beyond the scope of visual inspection
	\item Investigate the use of multivariate persistence for predicting properties of lattice QCD data 
\end{itemize}

\noindent
The remainder of this paper is structured as follows.  Section~\ref{sec::lattice_quantum_chromodynamics} introduces the relevant lattice QCD background required for these studies.  In Sections~\ref{sec::joint_contour_net} and~\ref{sec::topological_analysis} we review some of the topological techniques used to analyse scalar data, enabling us to pose a number of research questions specific to lattice QCD in Section~\ref{sec::research_question}.  In Section~\ref{sec::configuration_analysis} we present visual analysis of lattice QCD data.  From this we make observations that can be used to automate topological analysis for large ensembles of lattice QCD, as found in Section~\ref{sec::lattice_qcd_data_analysis}.  Finally, in Section~\ref{sec::conclusions} we reflect upon what physicists can gain by using multivariate topology to analyse their data and suggest future directions for the research.

\section{Lattice Quantum Chromodynamics}
\label{sec::lattice_quantum_chromodynamics}


Quantum Chromodynamics is the theory used to describe sub-nuclear interactions between quarks to form hadrons via the exchange of massless gluons.  Hadrons exist in two states; those consisting of three quarks are known as baryons, and quark-antiquark configurations make up the meson group.  Protons and neutrons are part of a subset of the baryon group called nucleons that combine with electrons to create the atoms that form the periodic table of elements.

\begin{figure*}[htb!]
	\centering
	\begin{minipage}[b]{\textwidth}
		\includegraphics[width=\textwidth]{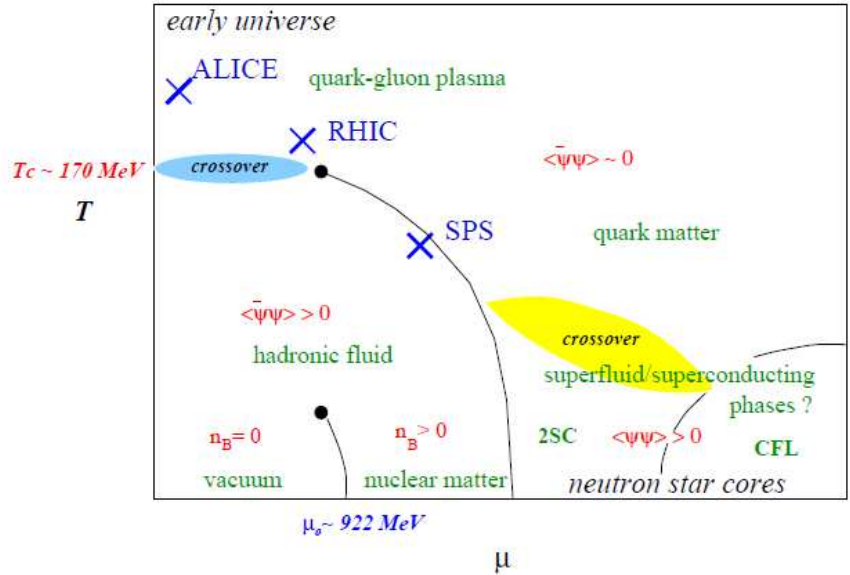}
		\caption{A proposed phase diagram for QCD by Hands~\cite{hands2001phase}.}
		\label{fig::phase_diagram}
	\end{minipage}
\end{figure*}

The behaviour of quarks and gluons across a range of temperatures and chemical potentials can be modelled using a phase diagram (Fig.~\ref{fig::phase_diagram}).  A proposed model given by Hands~\cite{hands2001phase} shows that situations at neutron star cores are most similar to a system with a non-zero chemical potential.  In addition, the diagram 
shows output from real-world experiments, allowing a sense of how work in this area fits in with experiments in nuclear and particle physics.  In order to test theories of the quark model computer simulations are employed to model quark-gluon interactions.  Effective computation of the strong forces described by QCD proceeds via a discretised model known as lattice QCD.



Lattice QCD rests on the Euclidean Path Integral formulation to strongly-interacting quantum field theory. As such, all quantities of physical interest are expressed as expectation values of products of field variables over a representative ensemble of configurations generated by Monte Carlo importance sampling techniques. Such expectations are called correlation functions;  they may relate fields at different spatial or temporal locations, yielding information such as the physical size and energy of strongly-bound particles called hadrons. Correlation functions may be specified either locally, or sometimes globally across the entire spacetime volume. In addition, for some observables UV-filtering techniques such as cooling are needed. As such QCD observables are multivariate in nature, and it is natural to apply multivariate analysis techniques in this domain.

\subsection{Lattice structure}

Kenneth Wilson was the first physicist to suggest that QCD could be approximated on a discrete lattice to model properties of quark and gluon fields \cite{WIL74}.  The structure of the lattice is a hyper-torus in Euclidean space-time, meaning that the three spatial dimensions and the time dimension are treated as equivalent.  Translational invariance $f(x) = f(x - L_{x})$ means that the lattice features periodic boundary conditions on every axis.  Temperature is defined as the inverse of the temporal lattice extent.


Quarks are placed onto the lattice at positions with integer indices, referred to as \emph{sites}.  In the remainder of this work we will label sites on the lattice using the notation $U(X)$ where $X \in \mathbb{Z}^4$.  These sites represent the starting point for various lattice field computations with a given origin.  From each lattice site four \emph{link variables} are used to model the gluon potential in the $x, y, z$, and $t$ directions between neighbouring sites.  Link variables are represented as $U_{\mu}(X)$, i.e. the variable defined on the link emerging from the site $X$ in the direction $\mu$.  When traversing the lattice, the relation  $U_{-\mu}(X)\equiv U_\mu^{\dagger}(X-\hat\mu)$ allows us to define movements in the reverse direction using the \emph{adjoint} form of link variables.

Each link variable is a member of the \emph{special unitary group} of matrices, identified using the notation $\su(n)$.  The value of $n$ represents the number of charge colours used in the gauge theory, with true QCD defined with $n=3$.  However, in this work we use a simplified \quotes{two colour} model of the theory using $\su(2)$ matrices.  One of the primary reasons for using a simplified model is that it allows us the freedom to vary the chemical potential of the system.

The \emph{cooling algorithm} is an established method of iteratively removing noise from lattice QCD configurations by modifying the values existing on the lattice.  However, the process is not without pitfalls; in particular, it is possible for over-cooling to remove the very features we intend to observe.  

\subsection{The Polyakov loop field}

\begin{figure}[htb]
	\centering
	\begin{minipage}[b]{.48\textwidth}
		\includegraphics[width=\textwidth]{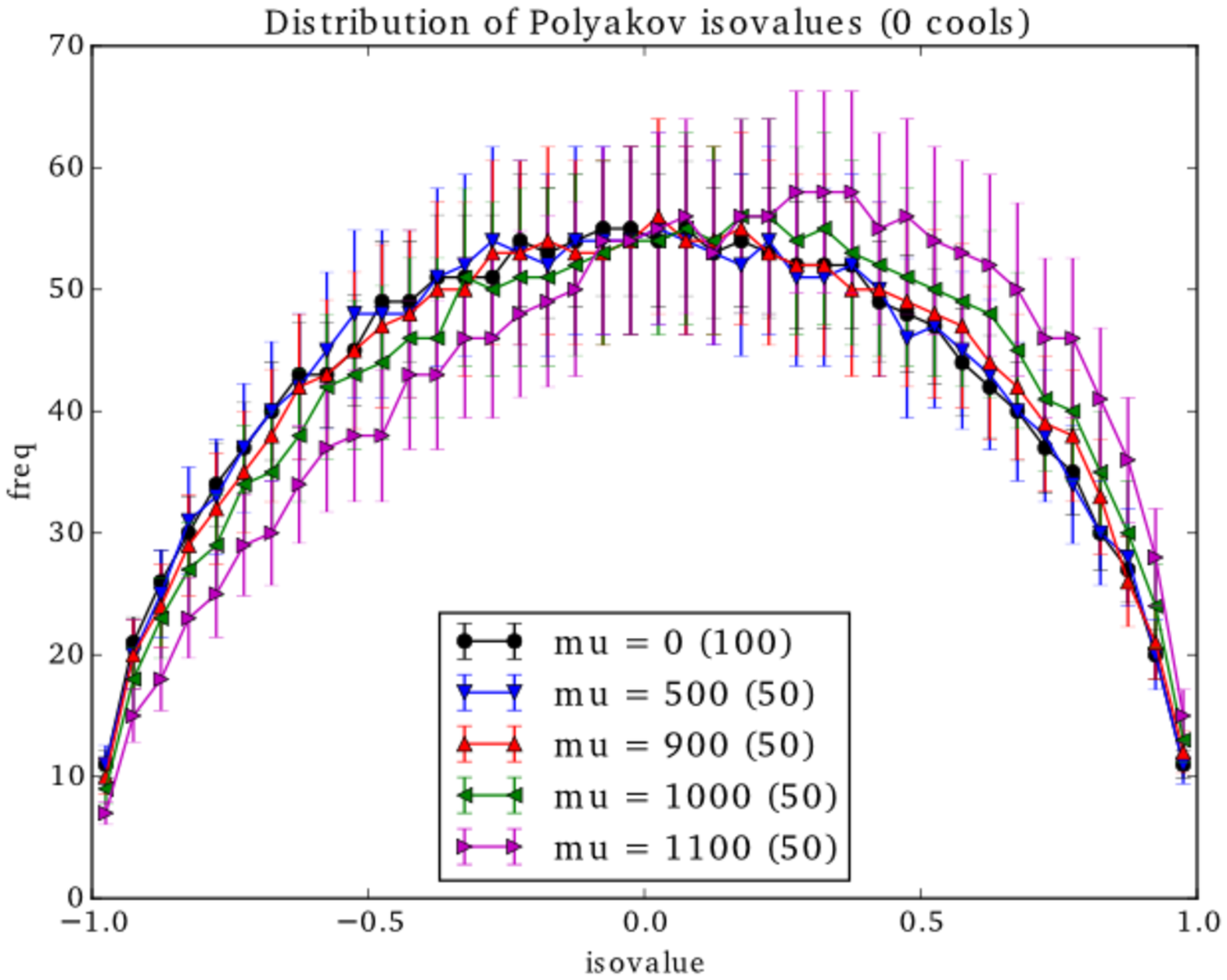}
		\caption{By binning the scalar values making up the Polyakov loop observable it is possible to see the breaking of symmetry at high chemical potentials ($\mu$) in un-cooled data.}
		\label{fig::polakov_scalar_distribution}
	\end{minipage}
	\hfill
	\begin{minipage}[b]{.48\textwidth}
		\includegraphics[width=\textwidth]{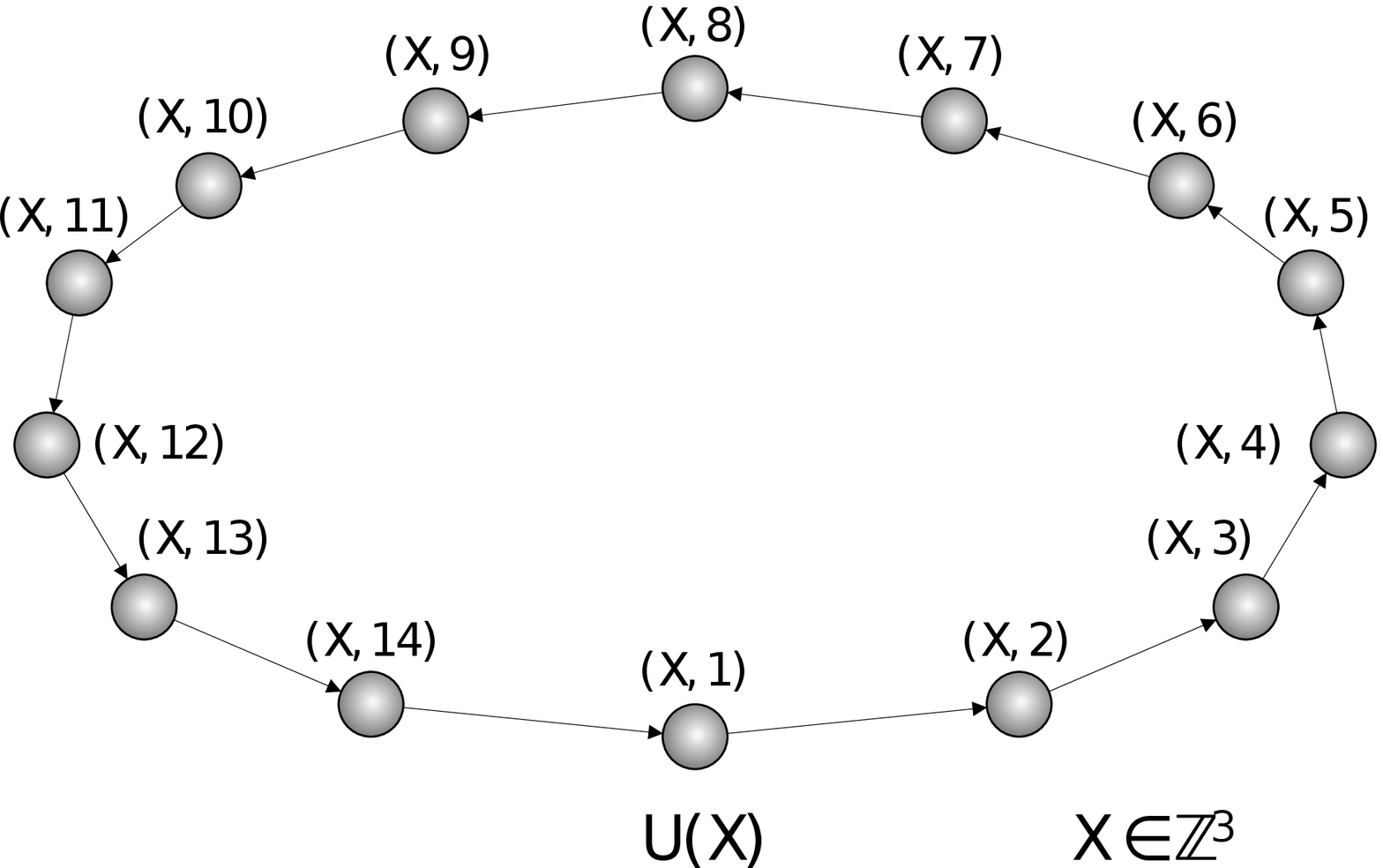}
		\caption{The Polyakov loop is a scalar field $f:\mathbb{R}^4 \mapsto \mathbb{R}^3$ computed by visiting each site in a given space-time direction.  In most cases we loop over the time axis.}
		\label{fig::polyakov_loop}
	\end{minipage}
\end{figure}

The Polyakov loop, otherwise known as the \emph{Wilson line} operator, is a lattice field that can be used as a method for computing the symmetry of a lattice.  Breaking of symmetry (Fig.~\ref{fig::polakov_scalar_distribution}) is one signal that can indicate a transition to a de-confined state, achieved by varying ensemble parameters such as temperature or chemical potential.  This method of locating critical temperatures is well established in $\su(2)$ and $\su(3)$ lattice gauge theories~\cite{mclerran1981monte, kuti1981monte, engels1981high, kajantie1981phase}.

From a computational point-of-view the Polyakov loop presents a convenient method for reducing the $\mathbb{R}^4$ lattice to a $\mathbb{R}^3$ scalar field.  In order to compute the Polyakov loop, we take the product of all time-like link variables from a given lattice site in three dimensions (see Eq.~\ref{eq::polyakov}).  On a lattice defined with a periodic time axis, as is the case in lattice QCD, the effect is a closed straight line, as visualised in Fig.~\ref{fig::polyakov_loop}.  Thus the Polyakov loop represents an attractive method for identifying de-confinement in a form that can easily be visualised and analysed using existing methods.

\begin{gather}
\label{eq::polyakov}
f(x) = \frac{1}{2}\Re(Tr(\prod\limits_{n=1}^{L_t}U_{\mu}(x + n\hat{\mu}))) \\
where\ \mu = 0, \ x \in \mathbb{Z}^3 \nonumber
\end{gather}

\section{Joint contour net}
\label{sec::joint_contour_net}

\begin{figure}
	\centering
	\includegraphics[width=\textwidth]{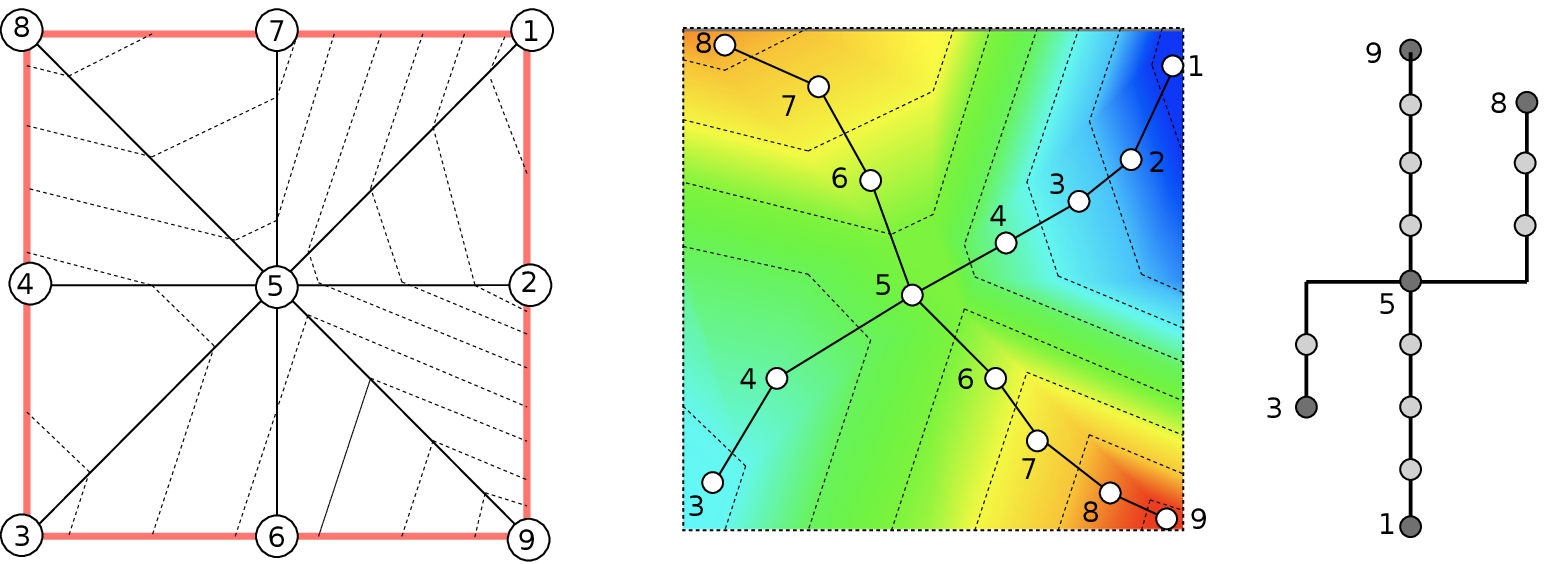}
	\includegraphics[width=\textwidth]{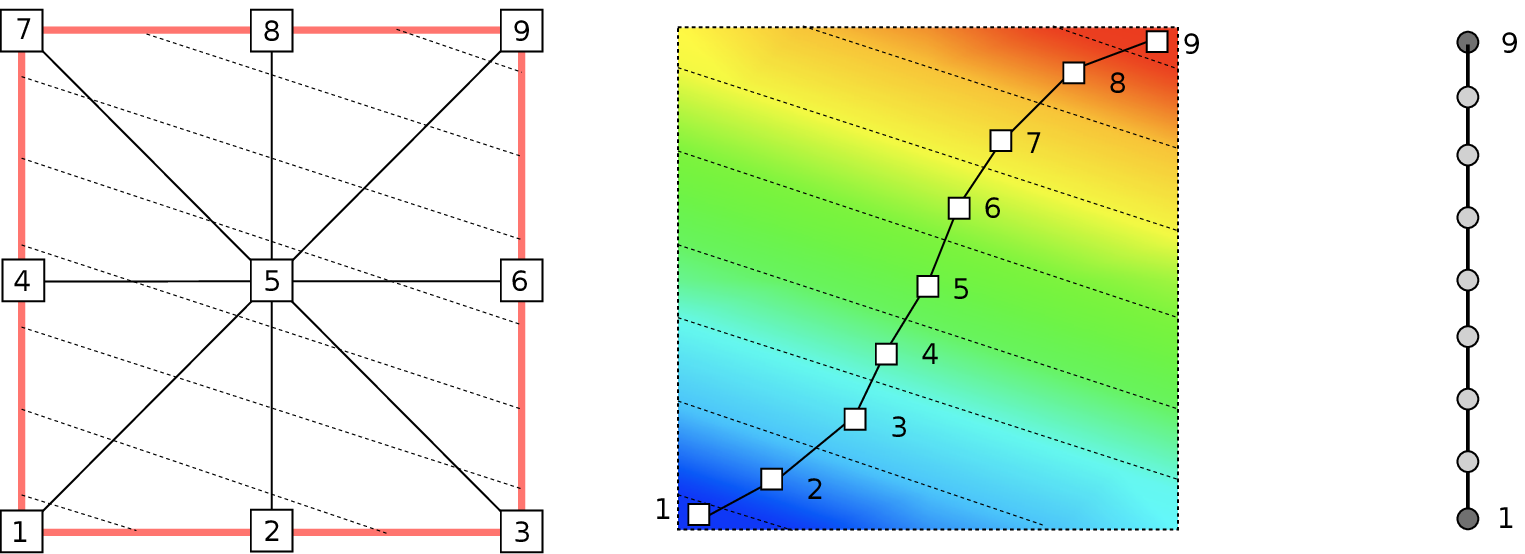}
	\caption{Two simple scalar functions defined on a simplicial grid (left) where the dotted	lines represent the quantisation intervals. The quantised contour tree for each function (right) is shown mapped to scalar field in the centre.}
	\label{fig::example_quantised_contour_trees}
	
	\includegraphics[width=.9\textwidth]{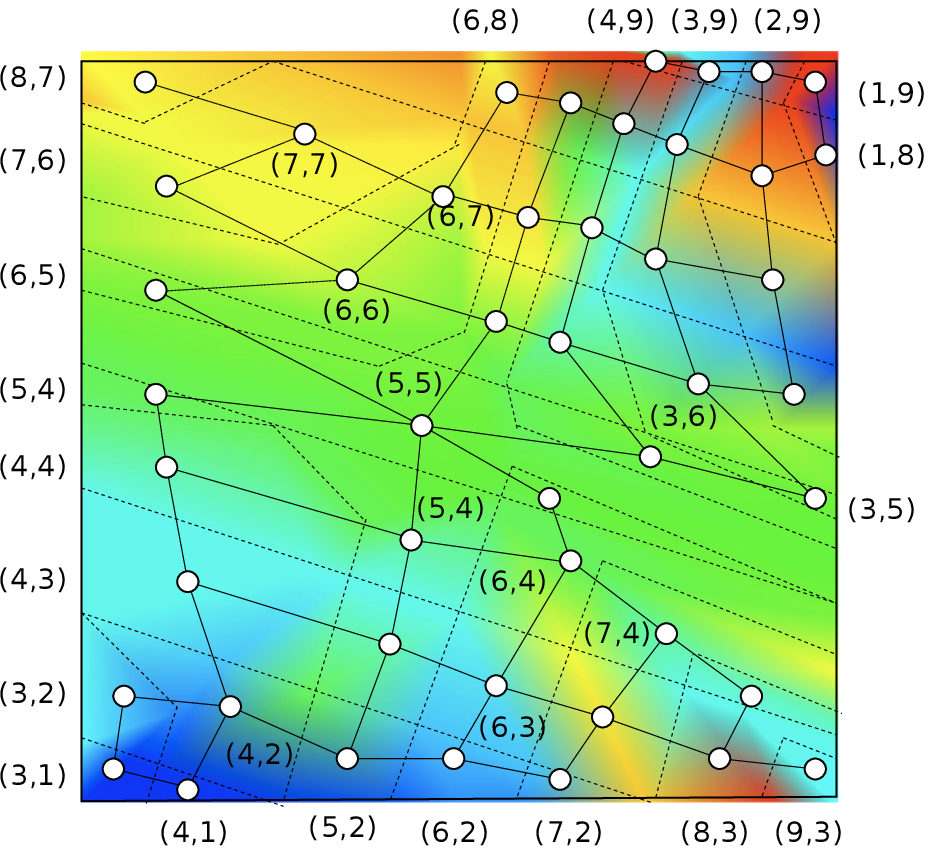}
	\caption{The JCN capturing the bivariate topology of the two simple functions shown in Fig.~\ref{fig::example_quantised_contour_trees}. The bivariate field is decomposed by overlaying the quantisation intervals of the two input fields (dotted line). A vertex is placed at the barycentre of each region, or joint contour slab, and edges mark adjacency.}
	\label{fig::example_jcn}
\end{figure}

The Reeb space is a generalisation of the Reeb graph~\cite{shinagawa1991constructing} capable of representing the topological structure of temporal or multivariate data sets~\cite{edelsbrunner2008reeb}.  Carr et al.~\cite{Carr2013} presented the first discrete representation of the Reeb space for functions of $n$ variables defined in an $\mathbb{R}^m$ dimensional space $f : \mathbb{R}^m \to \mathbb{R}^n$.  The JCN algorithm approximates the Reeb space as multivariate contours, or joint contour slabs, representing connected components of a joint level set, $f^{-1}(h) = \left\{ x \in M : round(f(x)) = h\right\}$, at isovalue $h \in \mathbb{Z}^n$. In situations where $n \geq m$ the JCN can be computed; however, the output is not an approximation of the Reeb space but instead a subdivision of the input geometry over $n$ variables. The JCN captures the Reeb space as an undirected graph structure, where vertices represents slabs of $n$ isovalue tuples, and edges are used to show adjacency between regions. An example JCN of two scalar functions is presented in Figures~\ref{fig::example_quantised_contour_trees} and~\ref{fig::example_jcn}.

\section{Topological Analysis}
\label{sec::topological_analysis}

Complicated datasets can be simplified using topological persistence measures in order to determine noise from features.  Carr et al.~\cite{carr2004simplifying} used this approach to assign basic measures of importance to the arcs of the contour tree.  Each arc represents the connected components of a single contour~\cite{carr2003path}; by following the path through the scalar field between the bounding vertices of an arc various properties of the contour can be computed.  Properties include the isovalue range, the approximated volume (number of samples), and the hypervolume (sample sum).  The contour tree is simplified by iteratively removing the least significant arc.  An alternative use for persistence measures is to relay information to the user regarding the properties of individual contours.  The contour spectrum~\cite{bajaj1997contour} used this approach to help users make informed about interesting features within scalar fields across a range of isovalues.  

For multivariate data persistence measures and simplification methods are more difficult to define.  One approach uses the Jacobi set \textemdash{} points where the gradient of multiple functions align or have a gradient of zero~\cite{edelsbrunner2004local}.  In situations where the multi-field represents temporal data this can be used to augment features with a lifetime parameter.  This approach was used in~\cite{bremer2007topological} to compute persistence in the context of the Morse-Smale complex.  However, when generalised to non-temporal functions defining persistence as a feature of the Jacobi set becomes a non-trivial task~\cite{suthambhara2011simplification}.  

The Reeb skeleton~\cite{chattopadhyay2015multivariate} collapses path connected regions of the JCN where the topology changes trivially to a single vertex.  Each vertex of the Reeb skeleton is augmented with persistence measures for the collapsed region including the number of JCN vertices and the approximated total volume of a merged region of slabs.  Using this approach lip pruning techniques, similar to those used in contour tree simplification~\cite{carr2004simplifying}, enable the progressive removal of noisy features in the multi-field.  More accurate measures of individual slabs can also be obtained by computing geometric properties directly from the meshes of the slabs.

In summary, persistence measures are typically used in two scenarios; the first is to simplify complex data by determine noise from features; the second is to provide additional information about distinct topological objects.  In this work we focus on the second approach as domain scientists have their own noise removal methods known as cooling.  We predict that as cooling progresses the function gradients of neighbouring cooling slices should see a gradual alignment of function gradients.  Hence, use of the Jacobi set could also provide a quantitative approach to predicting  optimal levels of cooling, helping to avoid \emph{over cooling} of the data.

\section{Research questions}
\label{sec::research_question}

Previous studies using the Reeb graph~\cite{thomas2016topological} suggest that the Polyakov loop observable may lose its ability to identify de-confinement under cooling.  We are interested to see if this phenomena can be identified using multivariate topology by comparing the Polyakov loop lattice observable at neighbouring cooling iterations.  From this it is hoped that the JCN can be used to provide additional information about lattice QCD simulations that are currently beyond the reach of domain scientists.  

We ask the following questions, focusing on the Polyakov loop under cooling:

\begin{itemize}
	\item Does the JCN allow us to question the relationship between objects with positive and negative isovalues in the Polyakov loop field?
	\item Can we use multivariate persistence measures to predict properties of the objects in the Polyakov loop field?
	\item By using the JCN can we learn anything regarding a lattice based noise reduction technique known as cooling?
\end{itemize}

\noindent
In order to answer these questions we first visually examine a single configuration at various levels of cooling in Section~\ref{sec::configuration_analysis}, we then repeat the process for multiple configurations in an in-depth ensemble study in Section~\ref{sec::lattice_qcd_data_analysis}.
\section{Visual analysis of lattice QCD data}
\label{sec::configuration_analysis}

\begin{figure}[!htb]
	\centering

	
	\begin{minipage}[b]{.10\textwidth}
		\includegraphics[width=\textwidth]{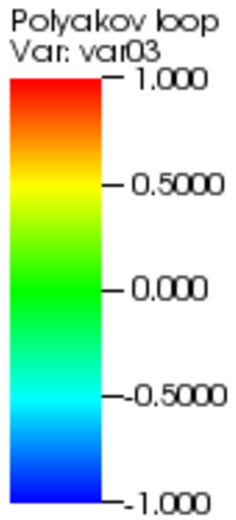}
	\end{minipage}
	\hfill
	\begin{minipage}[b]{.29\textwidth}
		\includegraphics[width=\textwidth]{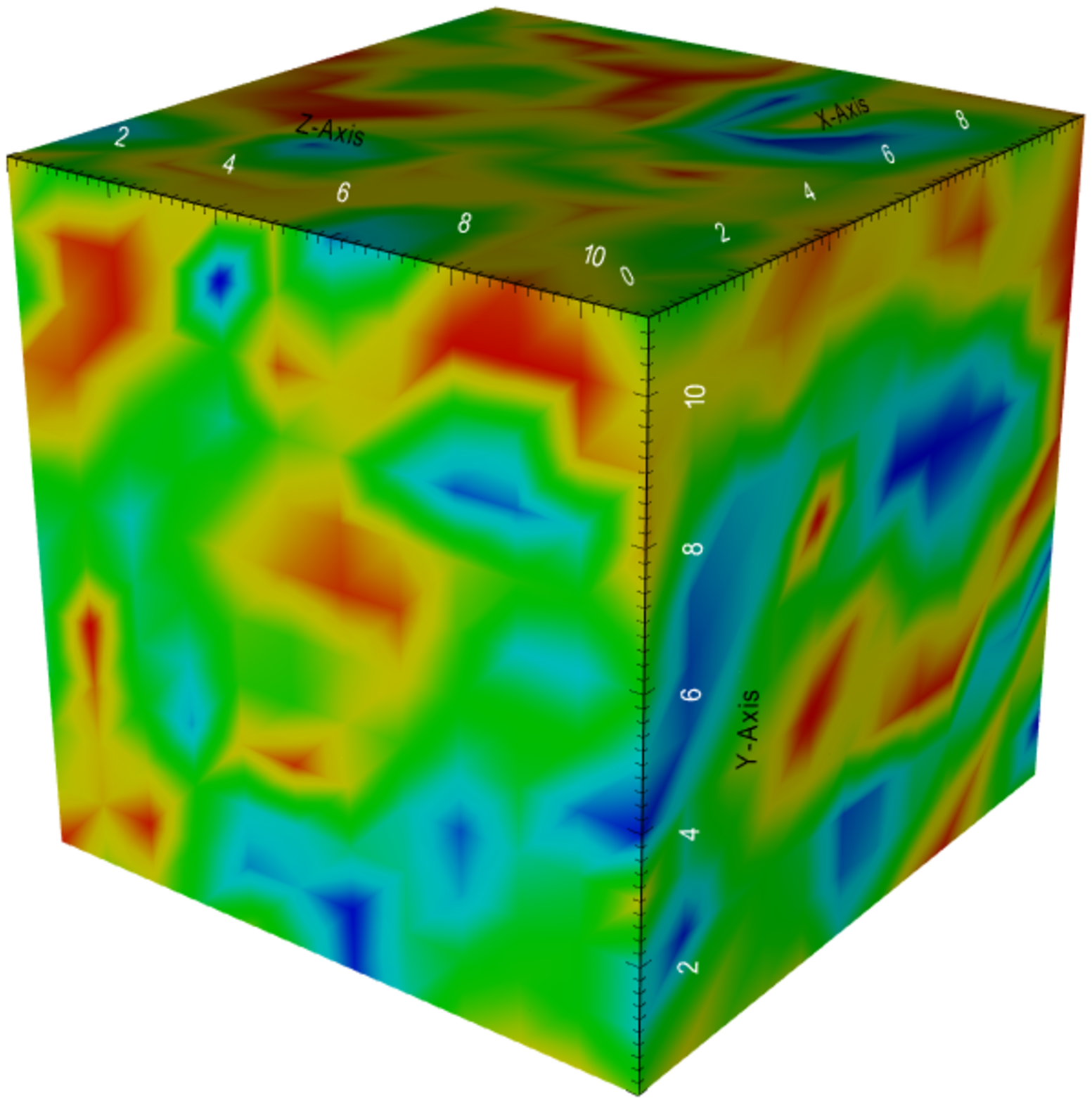}
	\end{minipage}
	\hfill
	\begin{minipage}[b]{.29\textwidth}
		\includegraphics[width=\textwidth]{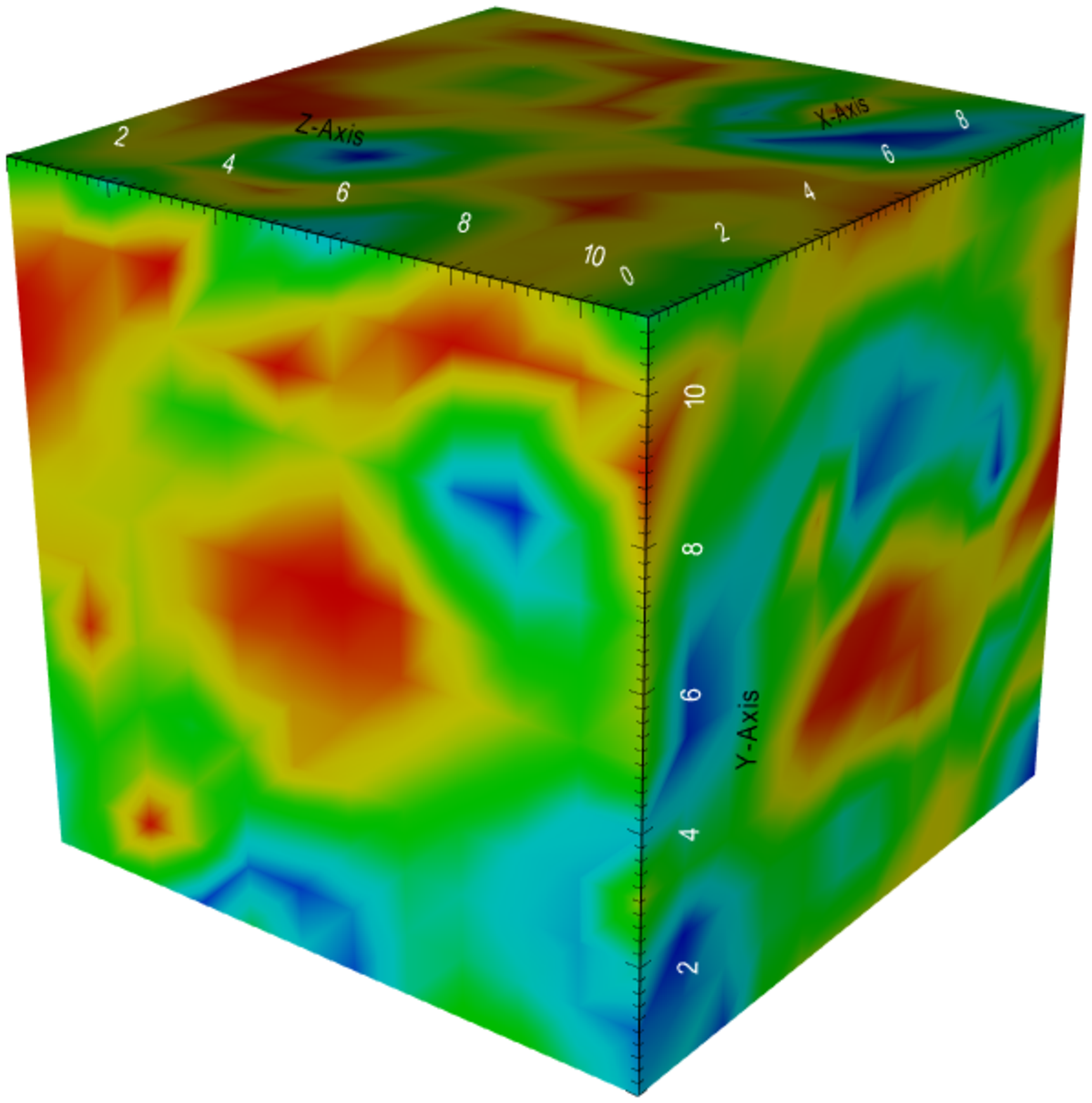}
	\end{minipage}
	\hfill
	\begin{minipage}[b]{.29\textwidth}
		\includegraphics[width=\textwidth]{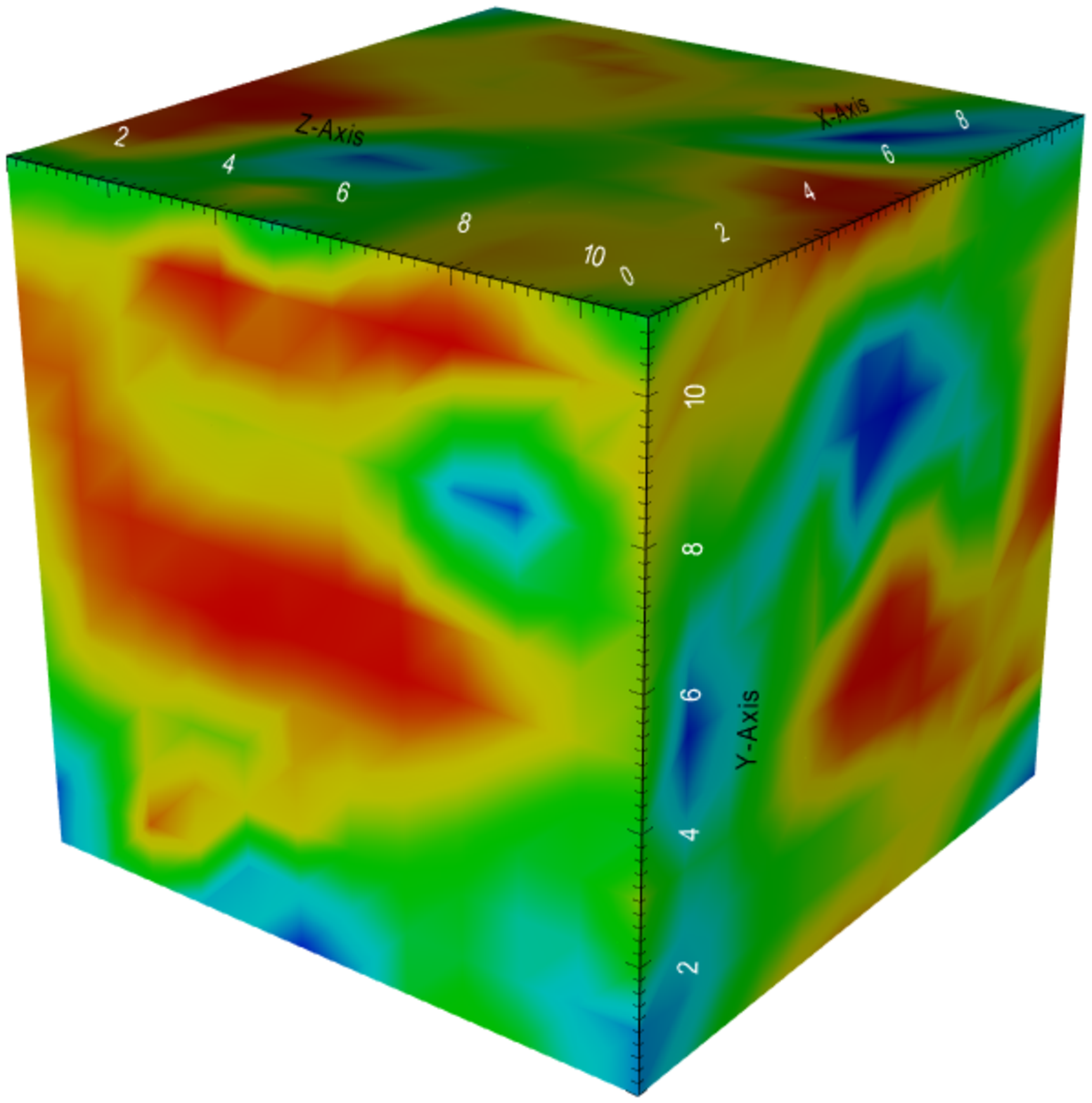}
	\end{minipage}
	\begin{minipage}[b]{.1\textwidth}
		~\\
	\end{minipage}
	\hfill
	\begin{minipage}[b]{.29\textwidth}
		\includegraphics[width=\textwidth]{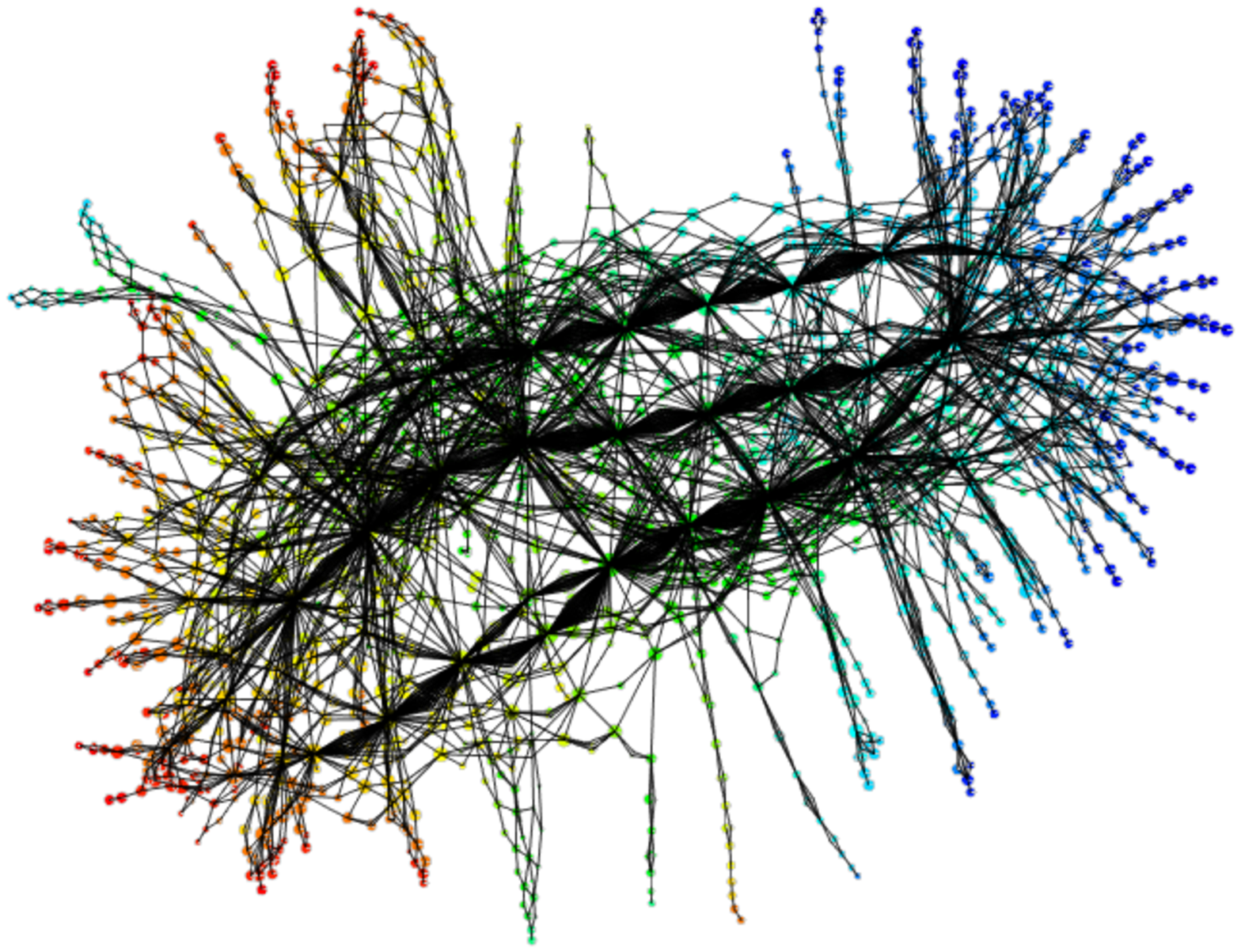}
	\end{minipage}
	\hfill
	\begin{minipage}[b]{.29\textwidth}
		\includegraphics[width=\textwidth]{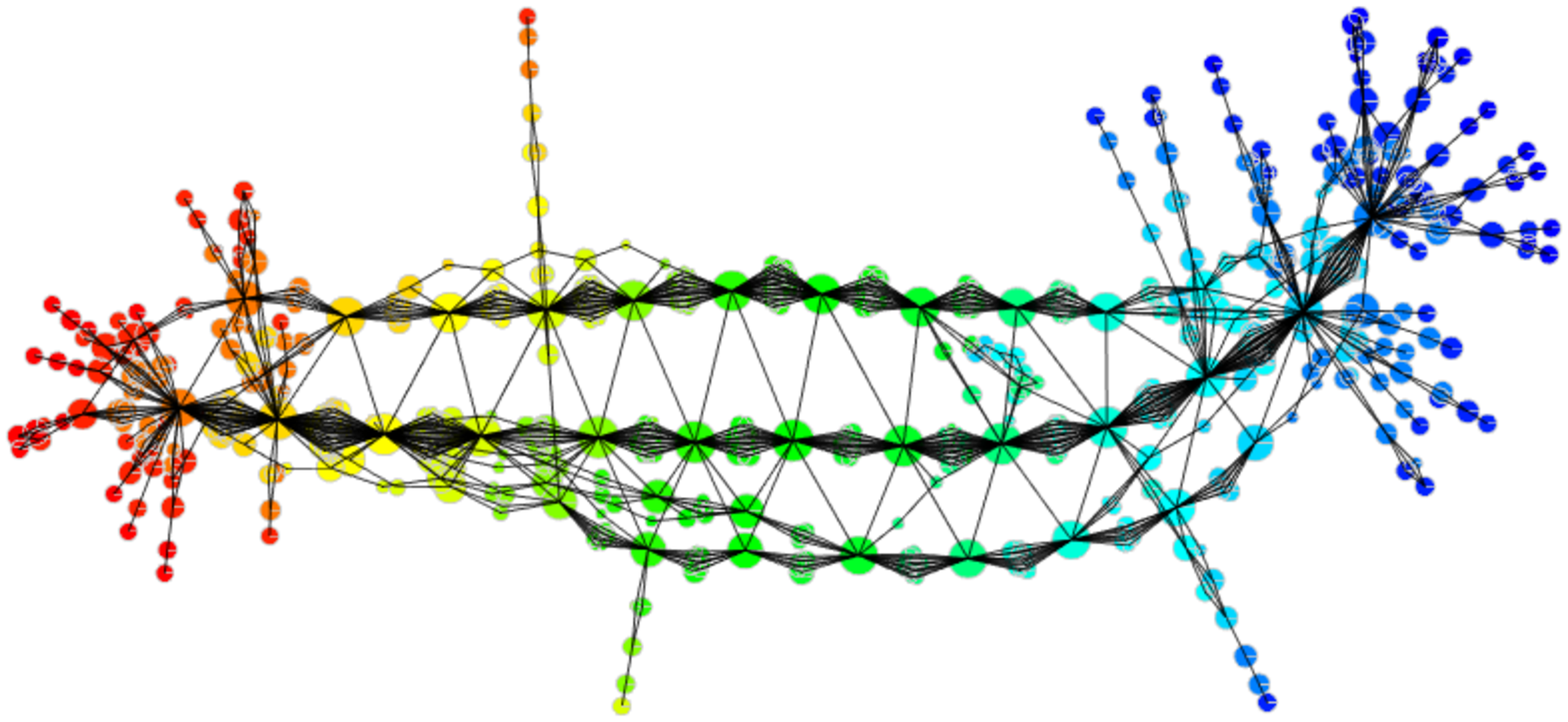}
	\end{minipage}
	\hfill
	\begin{minipage}[b]{.29\textwidth}
		\includegraphics[width=\textwidth]{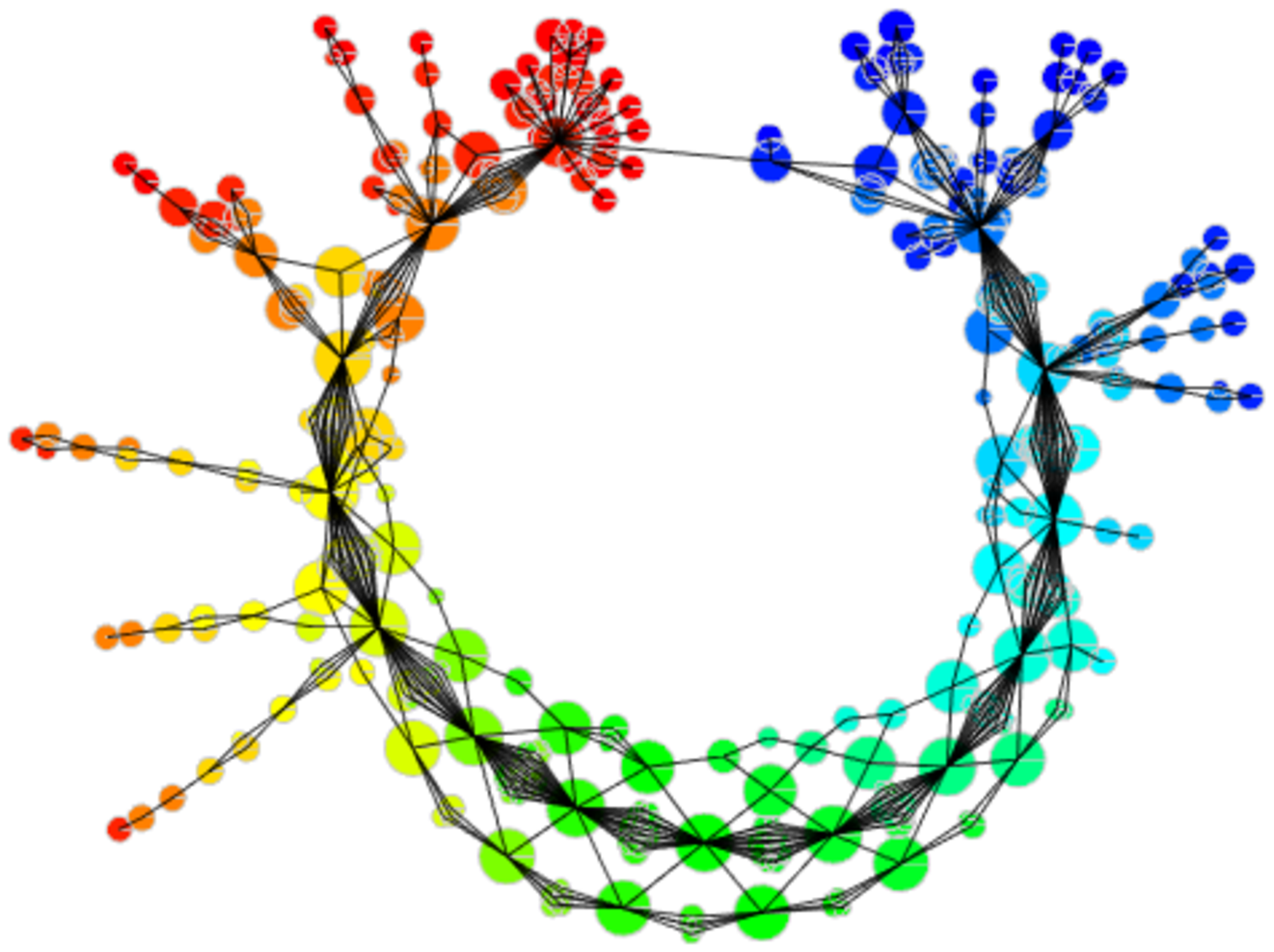}
	\end{minipage}
	\begin{minipage}[b]{.1\textwidth}
		~\\
	\end{minipage}
	\hfill
	\begin{minipage}[b]{.29\textwidth}
		\centering
		4 and 5 cools.
	\end{minipage}
	\hfill
	\begin{minipage}[b]{.29\textwidth}
		\centering
		9 and 10 cools.
	\end{minipage}
	\hfill
	\begin{minipage}[b]{.29\textwidth}
		\centering
		14 and 15 cools.
	\end{minipage}
	\caption{Visualising the slab geometry and joint contour net at cools 4 \& 5, 9 \& 10, and 14 \& 15.}
	\label{fig::slabs_and_jcn}
\end{figure}

The Polyakov loop data (Fig.~\ref{fig::slabs_and_jcn}) appears as blue regions of negative isovalues and red regions of positive isovalues separated by large regions of near-zero value in green.  Under cooling the slab geometry of the JCN reveals that these regions gradually merge into increasingly large topological objects.

The JCN captures this behaviour in a ladder-like structure.  Each vertex represents a joint contour slab associated with pair of isovalues, with adjacency shown by edges.  Individual objects in the field branch off the core structure at highly connected vertices with positive or negative isovalues.  Under cooling the ladder-like structure becomes increasingly simplified and objects tend to bunch up and branch off fewer vertices, each identifying a highly connected region of the topology.  For the purposes of automation and analysis it is desirable to summarise this information in a compact set of topological invariants.

An unexpected feature captured by the JCN net is the relationship between regions of the Polyakov loop field with positive or negative value.  At higher levels of cooling positive and negative regions become direct neighbours with no near-zero region in between.  The JCN indicates this by the edge between the blue and red vertices at 14 and 15 cools.

\section{Ensemble based analysis of lattice QCD data}
\label{sec::lattice_qcd_data_analysis}

%


Due to the quantum nature of lattice QCD each experiment must be computed multiple times as physical observables are defined as averages or \emph{expectation values}.  This results in an ensemble of suitably weighted configurations that are representative of the Feynman path integral over every possible state of the system.  The ensemble emerges from a Markov chain with updates generated by an algorithm, such as hybrid Monte Carlo~\cite{rothe2012lattice}, that respects a physical condition known as \emph{detailed balance}.


Therefore, each of these experiments are conducted as part of multi-ensemble studies~\cite{cotter2013towards}; in our particular work each ensemble is generated with a differing level of chemical potential ($\mu$).  This takes the form of computing around fifty separate JCNs for each ensemble; measures from the JCN are then computed and averaged to produce an ensemble average.  Typically domain scientists would then present this information with regard to chemical potential as a histogram or line graph to look for signals of phase changes in the simulated quark-gluon matter.

The experiments were carried out on a Dell cluster made up of four homogenous compute nodes each with access to 16GB RAM and a separate dedicated front-end node.  Each compute node contains four AMD Opteron 6376 CPUs, each with eight physical cores (16 logical cores) running 64-bit Debian 8.6.  The software is built using VTK 6.1~\cite{schroeder2006visualization} with extensions from the Multi-field Extension of Topological Analysis (META) project~\cite{meta2015} in C++11.  Quantisation parameters of the JCN were chosen to best suit the available resources in terms of memory and number of parallel processes.

\subsection{Results and analysis}

Below we present a selection of the most promising results gathered by analysing different aspects of the JCN in an ensemble setting.  Besides the two topological measures we discuss here we also analysed other aspects of the JCN graph structure.  We found that computing the ratio of Jacobi Nodes to JCN vertices gave an interesting overview of the effect of the cooling algorithm, with the ratio tending towards $1.0$ as the level of cooling reached its peak.  The ability to quantify convergence of the cooling algorithm presents a unique approach to avoid \emph{overcooling}, resulting in the unintended destruction of important lattice observables.

%
%

%

\runinhead{JCN vertex connectivity}  Connectivity within the JCN captures two interesting features.  Starburst effects, similar to those seen in~\cite{schunck2014description}, are captured by vertices with medium connectivity.  This represents sizeable regions of the Reeb space with two stable isovalues surrounded by multiple smaller regions where the isovalue is less stable.  Duke et al.~\cite{duke2012visualizing} suggest that these smaller regions are aliasing artefacts caused by the discrete nature of the algorithm.  

\begin{figure*}[htb!]
	\centering
	\begin{minipage}[b]{.98\textwidth}
		\includegraphics[width=\textwidth]{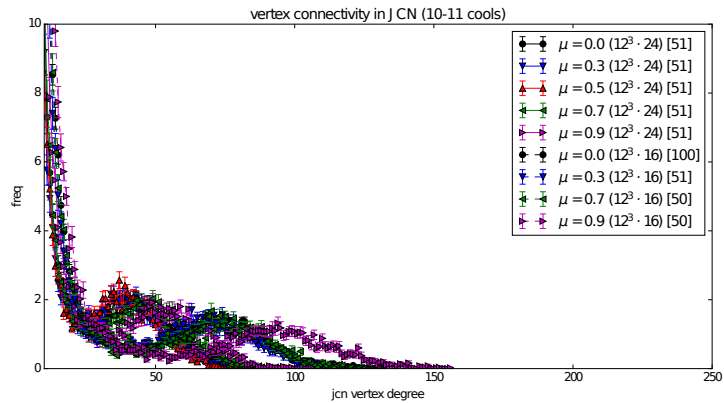}
		\caption{Degree of connectivity of vertices in the JCN as an average distribution for data cooled for 10 iterations.  A distinct banding effect can be observed, separating the hot and cold lattices and differing chemical potentials.}
		\label{fig::jcn_connectivity_10_11_cools}
	\end{minipage}
\end{figure*}

Also present in the data, particularly as the lattice is cooled, are a pair of vertices with very high connectivity at the extremes of the ladder like JCN structure (Fig.~\ref{fig::slabs_and_jcn}).  These correlate to stable regions with relatively large volumes with a number of smaller features branching from them.  Uncooled data show these branches throughout the isovalue range, but under cooling they gravitate to the largely connected vertices.

When considering the distributions for multiple ensembles we found interesting behaviour as the level of cooling was increased. For uncooled data there is very little difference between lattice sizes and levels of chemical potential ($\mu$), meaning all ensembles are roughly similar using this measurement.  However, as cooling takes effect we notice a distinct banding in the distributions (Fig.~\ref{fig::jcn_connectivity_10_11_cools}).  This appears to not only make a distinction between differing chemical potentials, but also the hot and cold lattice.  


\runinhead{Multivariate persistence measures}

\begin{figure*}[htb!]
	\centering
	\begin{minipage}[b]{0.44\textwidth}
		\includegraphics[width=\textwidth]{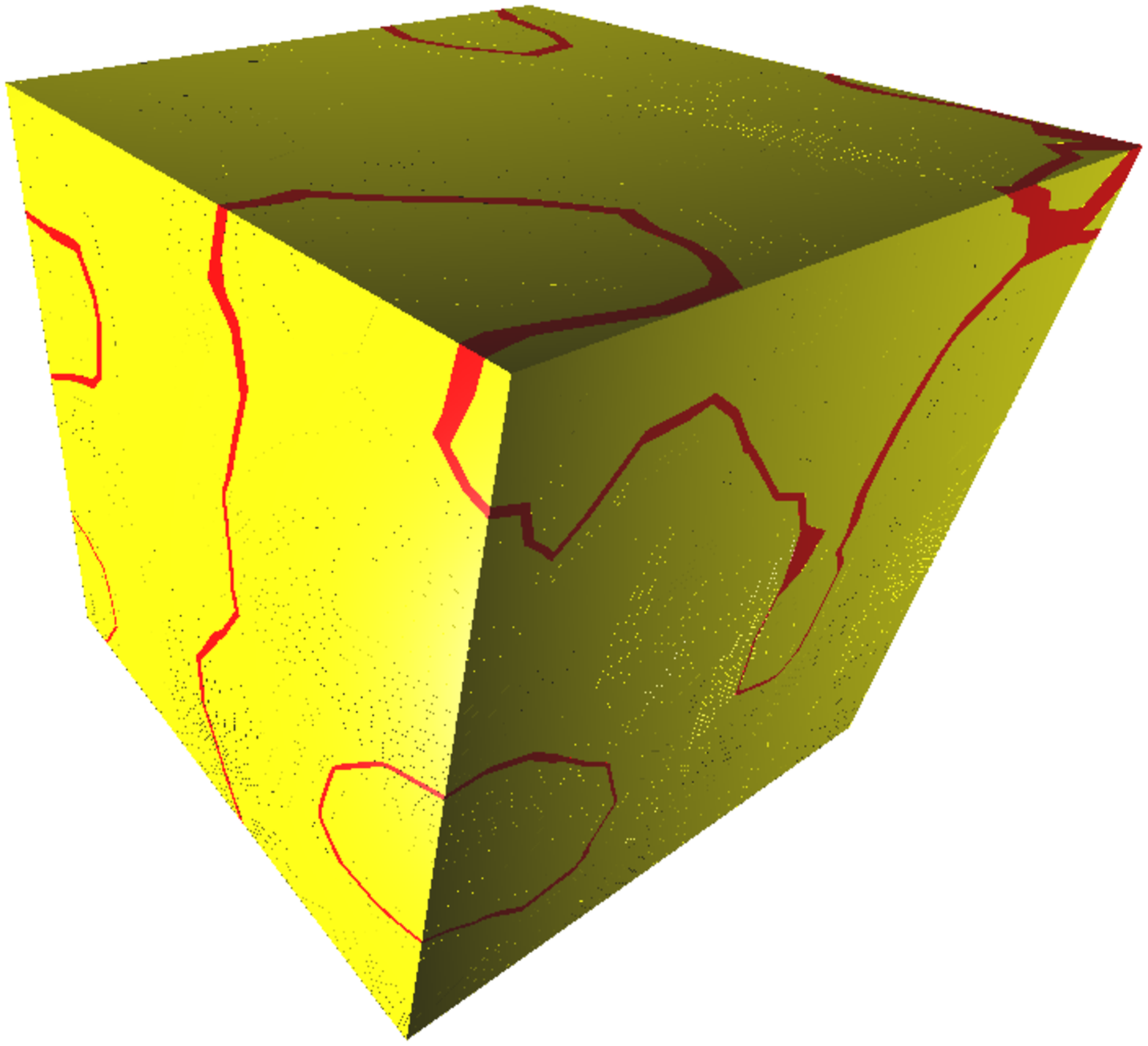}
	\end{minipage}
	\hfill
	\begin{minipage}[b]{0.44\textwidth}
		\includegraphics[width=\textwidth]{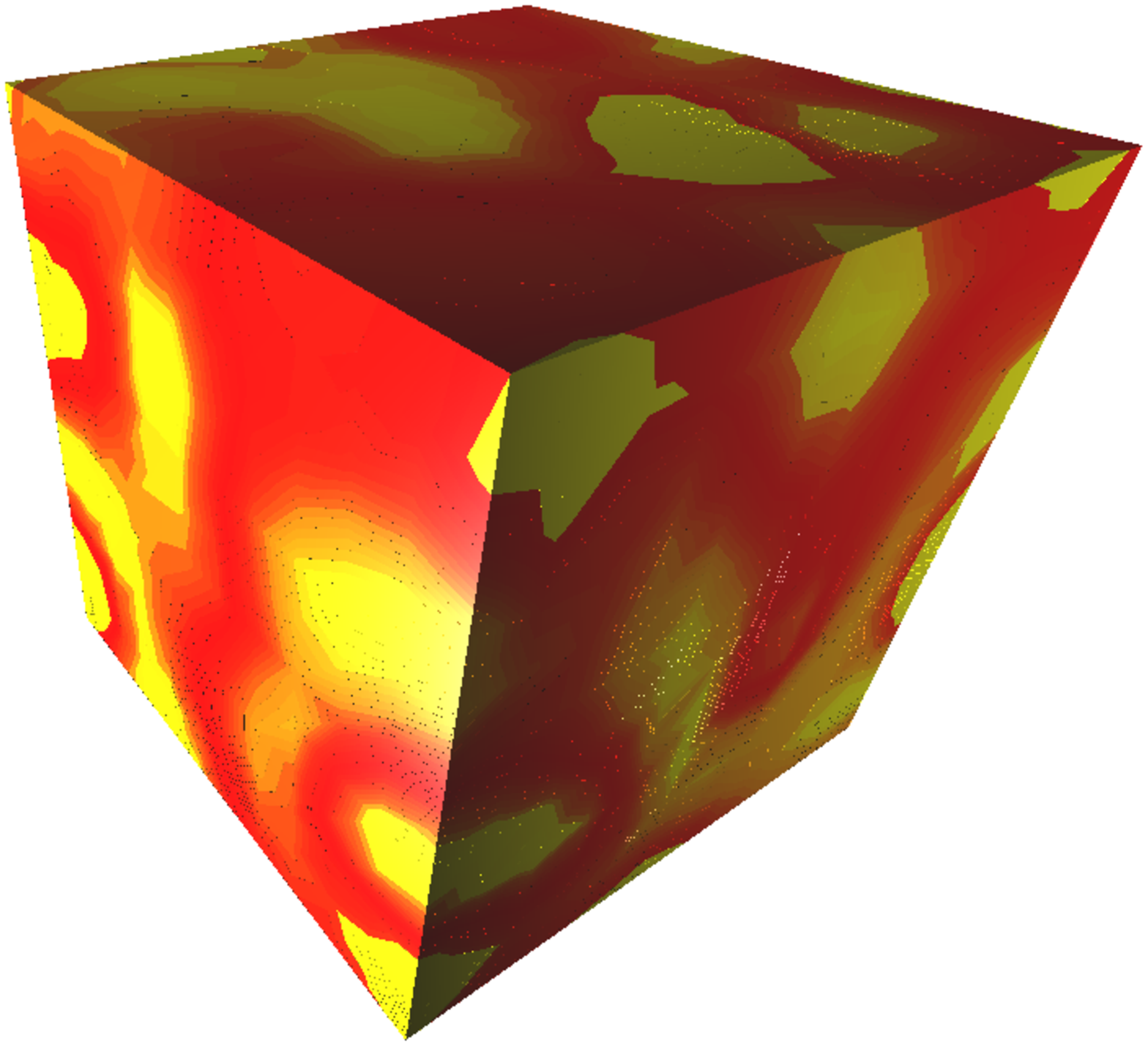}
	\end{minipage}
	\caption{A direct comparison of the multivariate persistence measures.  Left: measuring persistence as the number of triangles per slab.  Right: measuring persistence as the surface area of each slab.}
	\label{fig::polakov_slabs_persistence_comparison_mu0500}
\end{figure*}

Vertex connectivity gives a rough estimation of the number of stable regions present under cooling.  We next focus upon using persistence measures to calculate properties of these observables.  It is expected that highly connected vertices in the JCN correlate to regions of the lattice with non-trivial volumes indicating the presence of stable features present at both stages of the cooling algorithm.  Figure~\ref{fig::polakov_slabs_persistence_comparison_mu0500} demonstrates two persistence measures visually; working directly on the slab meshes, each of which is represented by a vertex in the JCN.  The triangle count measure is able to pick-out boundaries between regions in the multi-field, whereas the surface area highlights gives a gradient effect around prominent topological features.  

\begin{figure*}[htb!]
	\centering
	\begin{minipage}[b]{0.3\textwidth}
		\includegraphics[width=\textwidth]{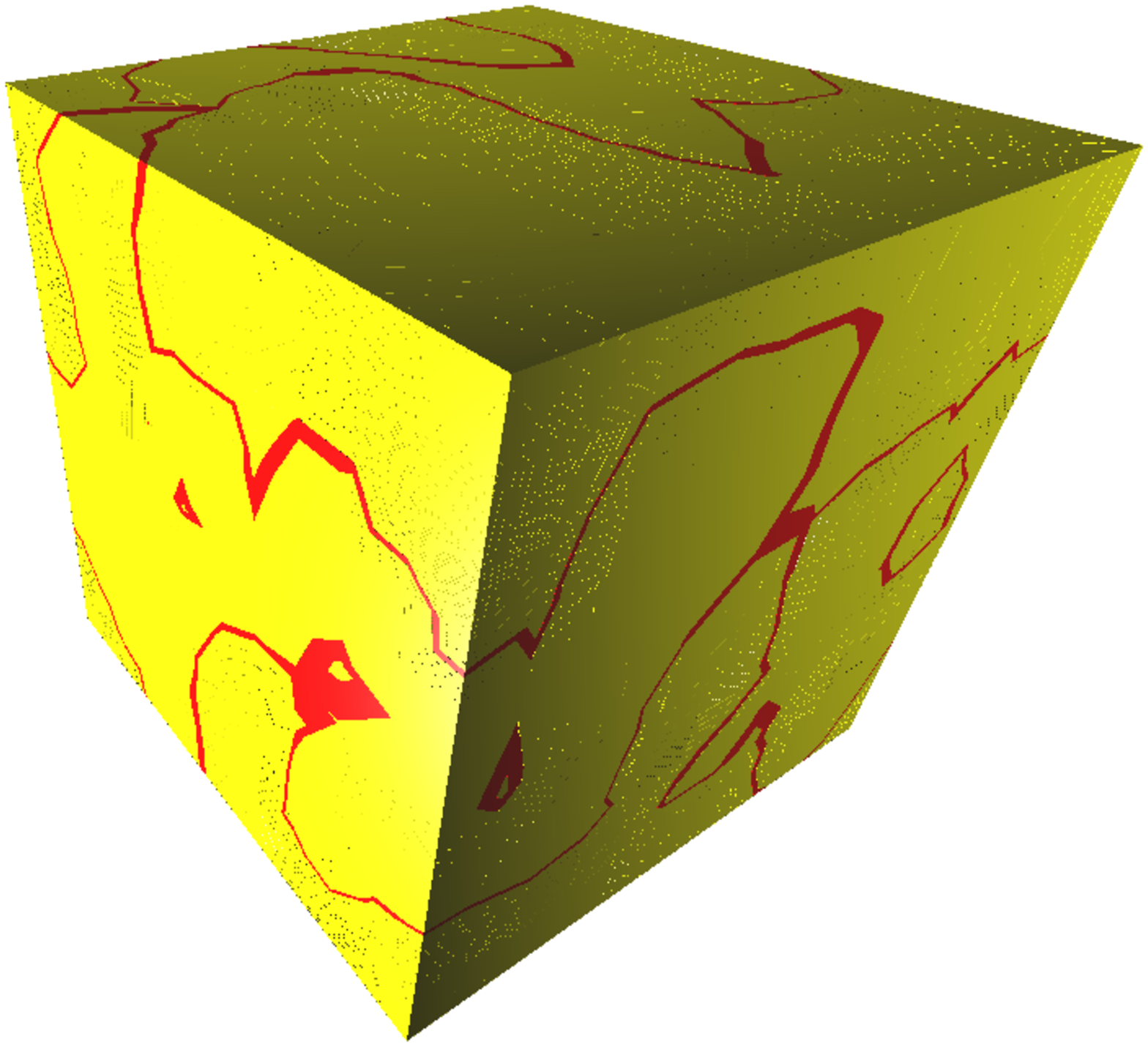}
	\end{minipage}
	\hfill
	\begin{minipage}[b]{0.3\textwidth}
		\includegraphics[width=\textwidth]{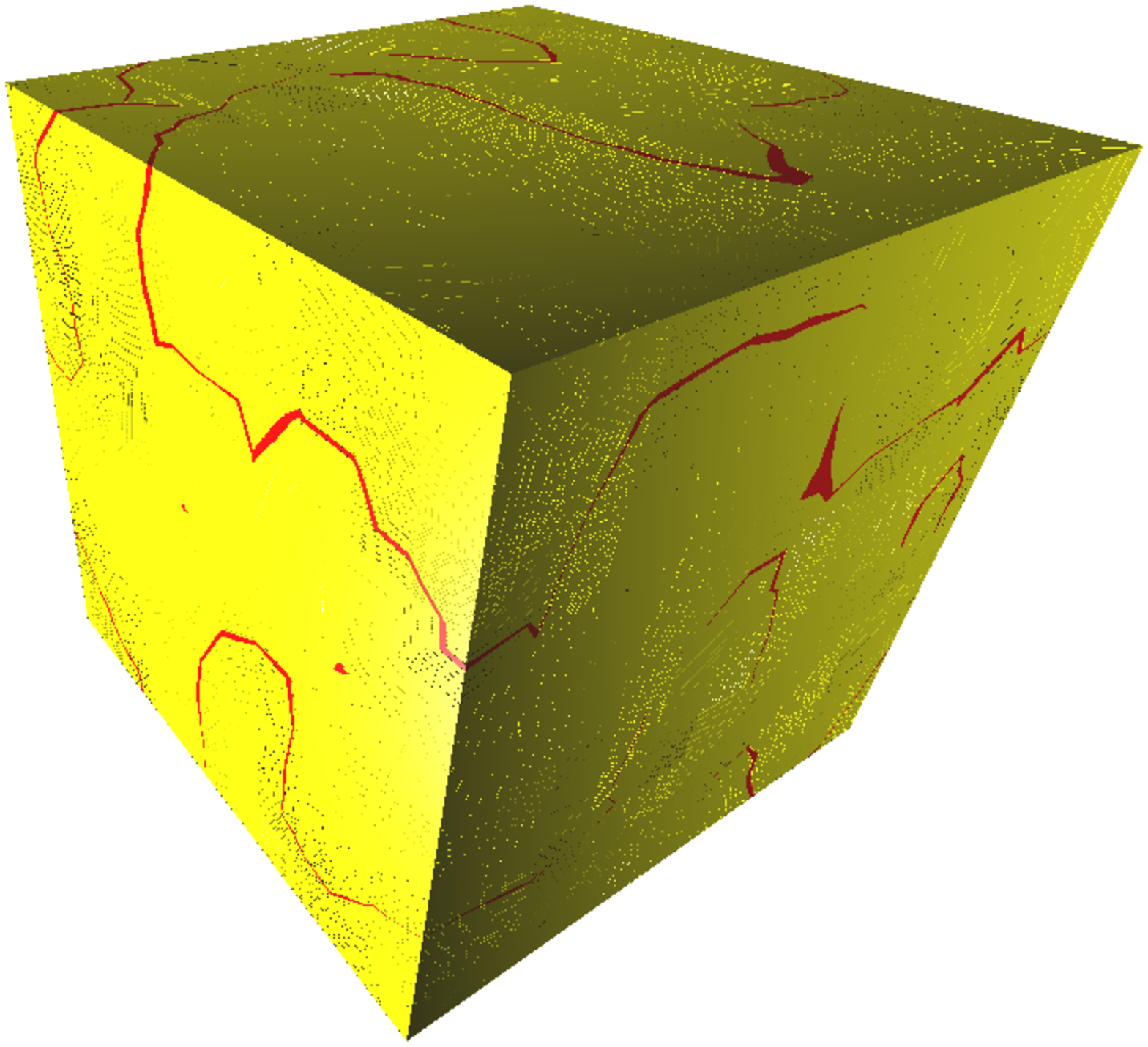}
	\end{minipage}
	\hfill
	\begin{minipage}[b]{0.3\textwidth}
		\includegraphics[width=\textwidth]{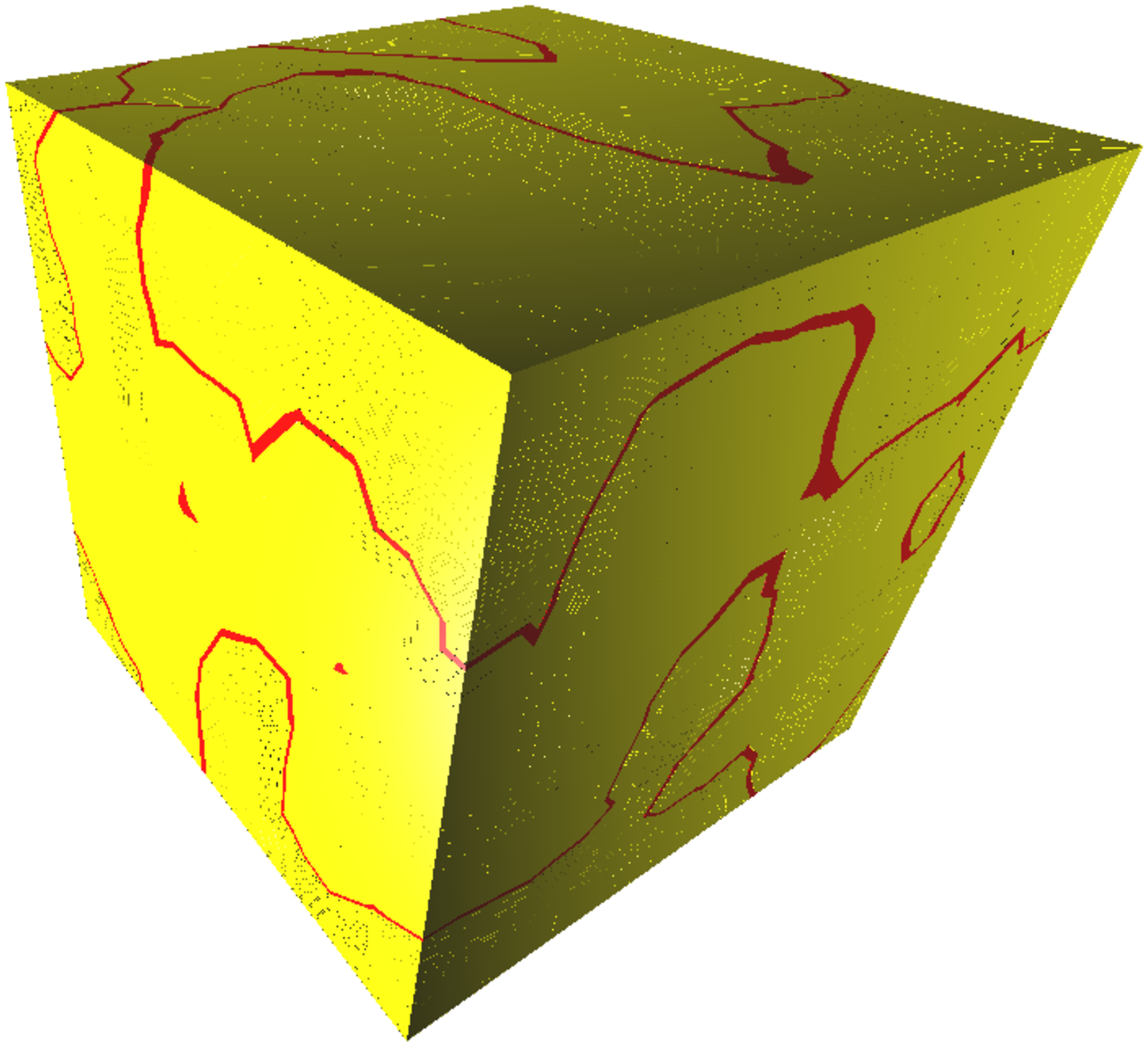}
	\end{minipage}
	\caption{Left: Quantised contours generated from Polyakov loop at 19 cooling iterations. Centre: Joint contour slabs are a union of the quantised contours at 19 and 20 cooling iterations. Right: Quantised contours generated from Polyakov loop at 20 cooling iterations.  Red regions represent most persistent slabs, measured by counting triangles.}
	\label{fig::polakov_slabs_triangles_mu0500}
\end{figure*}

In the case where a single variable is evaluated using the JCN the output is equivalent to a discrete segmentation of the volume by isovalue.  In Figure~\ref{fig::polakov_slabs_triangles_mu0500} we visualise this effect at two neighbouring levels of cooling alongside the bivariate topology of the same two inputs.  We use this approach to assist in interpreting features persisting at the two levels of cooling.

\begin{figure*}[htb!]
	\centering
	\begin{minipage}[b]{\textwidth}
		\includegraphics[width=\textwidth]{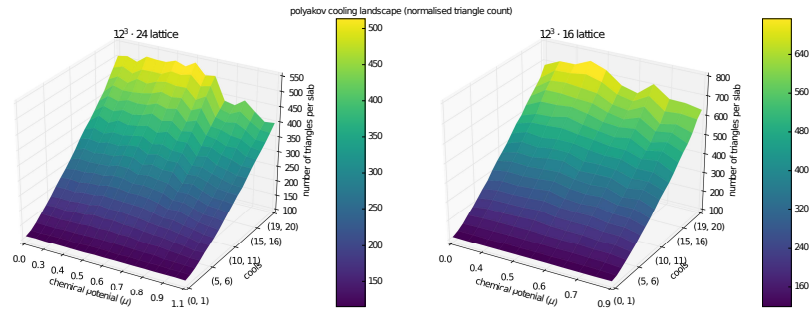}
		\caption{The average number of triangles per slab shown as the number of cooling iterations and chemical potential is varied.}
		\label{fig::polakov_landscape_triangle_count}
	\end{minipage}
\end{figure*}

To understand the effect of varying thermodynamic control parameters under cooling we computed persistence measures as the chemical potential ($\mu$) and number of cooling iterations is varied.  Persistence measures are averaged in each case by dividing by the total number of JCN vertices, relating to the number of slabs in the quantised Reeb space.  In Fig.~\ref{fig::polakov_landscape_triangle_count} we present this data as \emph{cooling landscape} generated by counting the average number of triangles per slab on the hot and cold lattices.  An upward trend is shown in the averages as the number of cooling iterations is increased and number of distinct topological objects decreases, representing the simplification effect the algorithm has on the input field and associated Reeb space.  A similar trend is visible when comparing with other persistence measures including the average surface area.

\begin{figure*}[htb!]
	\centering
	\begin{minipage}[b]{.96\textwidth}
		\includegraphics[width=\textwidth]{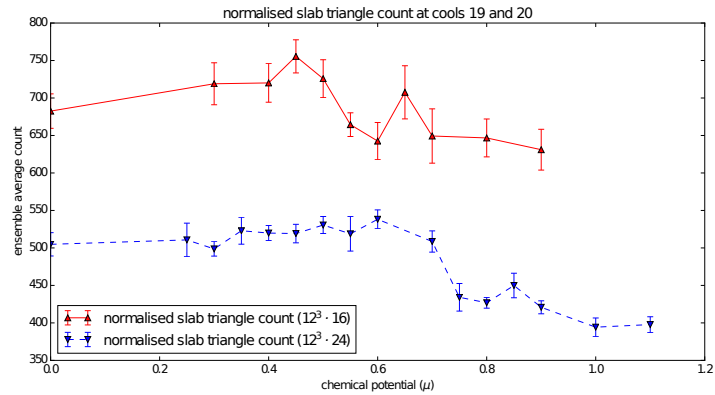}
		\caption{Average number of triangles per slab on the two lattices at cools 19 and 20.}
		\label{fig::polakov_triangle_count_19_20_cools}
	\end{minipage}
\end{figure*}

In Fig.~\ref{fig::polakov_triangle_count_19_20_cools} we show a direct comparison of the persistence as a measure of the number of triangles on the two lattices at the maximum number of cools.  Both lattices feature an initial largely flat region at low values of $\mu$, followed by a global peak.  A second peak follows higher in the chemical potential ($\mu$) range following a region with a downward trend in persistence before both distributions begin to plateau at high values of $\mu$.  Initial examination of the two distributions suggest that on the hotter lattice ($n_t=16$) the trend is shifted towards the lower end of chemical potentials.  This result seems to agree with recent evidence produced from statistical physics~\cite{boz2013phase} that see a shift in de-confinement on hotter lattices.  While this is an encouraging result, it should be noted that due to a limited number of configurations available for the analysis data exhibits large error bars. Consultation with physicists suggested that availability of a larger number of configurations will reduce the level of uncertainty.  The surface area of each also slab encouragingly converges to a similar distribution.  There are some subtle differences in the two persistence measures; most evident in the hotter ($n_t = 16$) lattice.  In particular there is a difference in the size of the peaks at $\mu = 0.45$ and $\mu = 0.65$.   

\section{Conclusions and future work}
\label{sec::conclusions}

We have presented a potential use of multivariate topology for analysing data from lattice QCD ensemble data sets.  The quantity of data required for analysis due to quantum mechanics means that visual inspection is not a feasible task.  Instead we have shown how a number of measures taken directly from the JCN can be displayed as domain specific parameters are varied.  Through the use of ensemble averages it is possible to understand if patterns present in the multivariate topology share a correlation with existing statistical physics predictions.  Current results suggest that some quantities, in particular multivariate persistence measures, could correlate well with physical observations.  We intend to publicly release all the results of this study in the future to allow other domain scientists to form their own interpretations the data.

In this work we have concentrated on a single lattice observable \textemdash{} the Polyakov loop.  However, lattice QCD presents many other observables that can be analysed for hints of de-confinement, many of which are defined as four dimensional space-time fields.  Future work is intended to focus upon analysing this data to look for correlations between other fields defined upon the lattice.

\begin{acknowledgement}
	This work used the resources of the DiRAC Facility jointly funded by STFC, the Large Facilities Capital Fund of BIS and Swansea University, and the DEISA Consortium (www.deisa.eu), funded through the EU FP7 project RI- 222919, for support within the DEISA Extreme Computing Initiative.  The work was also partly funded by EPSRC project: EP/M008959/1.
\end{acknowledgement}
%
%

%
%
\bibliographystyle{IEEEtran}
\bibliography{bibliography}

\end{document}